\documentclass[journal]{IEEEtran}
\usepackage{xcolor}
\usepackage{graphicx}
\usepackage{tikz}
\usepackage{color}
\usepackage{amsmath}
\usepackage{subfig}
\usepackage{algorithm}
\usepackage{algorithmic}
\usepackage{amssymb}
\usepackage{math}
\usepackage{caption}
\usepackage{array}

\begin{document}
\bibliographystyle{ieeetr}

\title{Blind MultiChannel Identification and Equalization for Dereverberation and Noise Reduction based on Convolutive Transfer Function
}
 \author{Xiaofei Li, Sharon Gannot and Radu Horaud
         
 \thanks{X. Li and R. Horaud are with INRIA Grenoble Rh\^one-Alpes, Montbonnot Saint-Martin, France. 
 E-mail: \texttt{first.last@inria.fr}
 }
 \thanks{Sharon Gannot is with Bar Ilan University, Faculty of Engineering, Israel. 
 E-mail: \texttt{Sharon.Gannot@biu.ac.il}
 }
 \thanks{This work was supported by the ERC Advanced Grant VHIA \#340113.}
 }

\maketitle

\begin{abstract}
This paper addresses the problems of blind channel identification and multichannel equalization for speech dereverberation and noise reduction. 
The time-domain cross-relation method is not suitable for blind room impulse response identification, due to the near-common zeros of the long impulse responses. 
We extend the cross-relation method to the short-time Fourier transform (STFT) domain, in which the time-domain impulse responses are approximately represented by the convolutive transfer functions (CTFs) with much less coefficients.
The CTFs suffer from the common zeros caused by the oversampled STFT. We propose to identify CTFs based on the STFT with the oversampled signals and the critical sampled CTFs, which is a good compromise between the frequency aliasing of the signals and the common zeros problem of CTFs. 
In addition, a normalization of the CTFs is proposed to remove the gain ambiguity across sub-bands. 
In the STFT domain, the identified CTFs is used for multichannel equalization, in which the sparsity of speech signals is exploited. We propose to perform inverse filtering by minimizing the $\ell_1$-norm of the source signal 
with the relaxed $\ell_2$-norm fitting error between the micophone signals and the convolution of the estimated source signal and the  CTFs used as a constraint. This method is advantageous in that the noise can be reduced by relaxing the $\ell_2$-norm to a tolerance corresponding to the noise power, and the tolerance can be automatically set. 
The experiments confirm the efficiency of the proposed method even under conditions with high reverberation levels and intense noise. 
\end{abstract}


\section{Introduction}
\label{sec:introduction}

The goal of speech dereverberation is to remove the reverberations from the received microphone signals to improve the speech intelligibility for both human listening and machine recognition. 
The output of a dereverberation system could be an estimation of either the anechoic source signal or the source signal with some early reflections, since early reflections do not deteriorate the speech intelligibility \cite{arweiler2011}. 
Multi-channel dereverberation methods are generally more powerful than single-channel methods, due to the use of the extra spatial information and the relation across channels. 

Multichannel dereverberation broadly includes the following techniques: (i) the spectral enhancement technique removes reverberation by spectral subtraction. Many techniques have been proposed to estimate the power spectral density of reverberation, such as statistical model \cite{habets2009}, coherent-to-diffuse power ratio \cite{schwarz2015};
(ii) the linear prediction technique is used in the weighted prediction error (WPE) algorithm \cite{nakatani2008,yoshioka2011,jukic2015}: reverberation is first estimated by filtering the microphone signal with the prediction filter, and then subtracted from the microphone signal (in a way, the prediction filter could be considered as the inverse of the channel filter, which is estimated by maximizing a likelihood function derived from a \emph{time-varying all-pole} model of source spectra);
(iii) probabilistic techniques apply dereverberation by maximizing the likelihood of the generative model of the microphone signals, such as in \cite{ito2014,schwartz2016}, and
(iv) the multichannel equalization technique first blindly identifies the channel filters, then applies the inverse filtering on the microphone signals. 

The focus of this paper is a multichannel equalization technique. For a single-input multiple-output (SIMO) system, the blind channel identification can be performed based on the second-order statistics, 
such as the subspace method \cite{moulines1995} or the cross-relation method \cite{xu1995}. The cross-relation method identifies the channel filters by detecting the eigenvector corresponding to the unique zero eigenvalue of the covariance matrix of microphone signals. 
A noise subspace method proposed in \cite{gannot2003} exploited the multiple eigenvectors corresponding to the zero eigenvalues of the over-modeled covariance matrix. 
These noise-subspace based methods, especially the cross-relation method, are vulnerable to noise interference and the filter length determination error. Thence they require prior knowledge of the exact filter length. 
However, in acoustic dereverberation, the room impulse response (RIR) is a time sequence with the variance exponentially asymptotically approaching to zero. 
The filter length is not measureable due to the long tails, in other words, the truncation point is difficult to determine.  
For the case of small tails, \cite{liavas1999} proposed the channel under-modeling method that only considers the significant part of the filters, 
and a rank detection method was proposed in \cite{liavas1999blind} to determine the length of the channel filter corresponding to the significant part. 
However, these methods are applicable only when a noticeable gap exists between the significant part and the small tails, which is obviously not the case of RIR.  
Based on the least mean squares method, the frequency-domain adaptive cross-relation method was proposed in \cite{huang2003}, and was applied to speech separation and dereverberation in \cite{huang2005}. 
One of the identifiability conditions of the second-order statistics-based method is that the multiple channels are co-prime, namely they do not share any common zeros. 
It is shown in \cite{lin2006} that a large number of near-common zeros exists for the long channel filters, which will deteriorate the performance of channel identification. 
A forced spectral diversity algorithm \cite{naylor2008,lin2012} was proposed to mitigate the near-common zeros problem. In this method, the distance between the zeros of the channels is enlarged by filtering one channel with the spectral shaping filters, and then under-modeling this channel.  
Overall, the channel identification techniques mentioned above are hardly applicable to real-world RIR identification due to the filter length indetermination  and the near-common zeros problem.
As a result, experiments based on  these methods are only carried out with channel filters that (i) do not have small tails, either the simulated filters or the RIRs with the tails being cut off, and (ii) are very short, 
e.g. from several taps to a few hundreds taps, which is not the case of RIRs that usually have thousands of taps. 

For multichannel equalization using the known channel filters, ideally, the classical multiple-input/output inverse theorem (MINT) method can perfectly recover the source signal. 
In MINT, the inverse filters of the channel filters are obtained by using the least square method. However, MINT is sensitive to the filter perturbations and the additive noise in the microphone signals. 
To improve the robustness of MINT to the filter perturbations, many techniques have been proposed by preserving not only the direct-path impulse response but also the early reflections, 
such as channel shortening \cite{kallinger2006}, infinity- and $p$-norm optimization-based channel shortening/reshaping \cite{mertins2010}, partial MINT \cite{kodrasi2013}, and relaxed multichannel least squares \cite{lim2014}.
To improve the robustness of MINT to the noise interference, a filter energy regularization was adopted in \cite{kodrasi2013,hikichi2007}, since an inverse filter with small energy would avoid the amplification of noise.
To jointly perform dereverberation and noise reduction, the regularization of the output noise power was proposed in \cite{kodrasi2016} in the framework of multichannel equalization. 
The techniques mentioned above first estimate an inverse filter, being independent to the signals, and then apply the inverse filtering to the microphone signals.  

This work aims to develop a blind multichannel equalization method for speech dereverberation in realistic scenarios. We propose to perform blind channel identification and multichannel equalization in the short-time Fourier transform (STFT) domain. 
The channel filters in each frequency band are much shorter than the time-domain filters, consequently they have much less near-common zeros. Moveover, the sparsity of speech signal in the STFT-domain 
is desirable for recovering the source signal, especially for the noisy case. It is known that STFT can be interpreted as the multi-rate filter banks \cite{vetterli1987}. 
The adaptive filtering in sub-bands has been widely studied \cite{gilloire1992,reilly2002}, and applied to acoustic echo cancellation. While the sub-band blind channel identification has been rarely studied. 
In \cite{gannot2003}, the noise subspace method was executed in sub-bands, however it was not applied to real scenarios. 
The channel filters are identified to a different gain factor for different sub-bands, which causes the gain ambiguity \cite{gannot2003}. A gain ambiguity correction method was proposed in \cite{castro2010} based on the overlapping pass-band regions. 
The sparsity of speech in the STFT domain has been exploited in \cite{kodrasi2016robust} by adding the signal sparsity function to the MINT function of time-domain filter as a regularization. 
However, the MINT function is heterogeneous with the regularization term, which makes it difficult to automatically set the regularization factor.

This paper proposes a blind convolutive transfer function (CTF) identification method, and a CTF-based inverse filtering method for dereverberation and noise reduction. 
In the STFT domain, the time-domain channel filters can be represented as the cross-band filter \cite{avargel2007}. 
To simplify the analysis in practice, we consider the use of the CTF approximation, i.e. only the band-to-band filter, as used in \cite{talmon2009}. 
Regarding the CTF indentification, this paper has the following contributions.
First, the influence of the STFT configuration on the signal reconstruction, CTF approximation and common zeros issue are analyzed in \ref{ssec:stft} in detail. 
Briefly stated, in each frequency band, the cross-relation method \cite{xu1995} is extended for CTF identification, which is not trivial. 
CTF only takes the band-to-band filter, while it disregards the cross-band filters, thus suffers from an under-modeling  error.  
The frequency response of the CTF will not be fully activated, i.e. zeros exist, for the oversampling case, namely the frame step is less than the window length.
The sub-band filter is common to all channels, thus the zeros are also common to all channels, which is problematic for the cross-relation method. 
Conversely, critical sampling will lead to the severe frequency aliasing, and thus to poor signal reconstruction. 
Second, to achieve a good compromise, we propose a novel configuration, briefly, the signals are oversampled to avoid the frequency aliasing, and the channel filters are forced to be criticaly sampled to avoid the common zeros problem.
Third, beyond exploiting the eigenvector corresponding to the unique zero eigenvalue, we propose to estimate the channel filters by solving a constrained least-squares problem, which is robust to the noise interference and the filter length determination error.
Fourth, to remove the gain ambiguity accross sub-bands, we propose to use the channel filters normalized by the first coefficient of the first channel, as a result, the time-domain equalized signal 
is the original source signal filtered by the early (corresponding to the length of STFT window) impulse response of the first channel. In other words, the dereverberated signal preserves the early reflections.   

As for the inverse filtering method based on identified CTFs, this paper has the following contributions. First, an optimization problem is proposed for both dereverberation and noise reduction. 
In detail, we propose to recover the source signal by exploiting the $\ell_2$ fit between the microphone signals and the channel filtered source signal, rather than explicitly estimating the inverse filters. 
The sparsity of the estimated signal is exploited by minimizing its $\ell_1$ norm. To reduce the noise caused by the filter perturbations and the microphone noise interference, 
the $\ell_2$ fitting error is relaxed to a tolerance corresponding to the noise power. Overall, the source signal is obtained by solving an $\ell_1$ minimization problem with an $\ell_2$ fit constraint. 
Second, the convex optimization algorithm primal-dual interior-point method \cite{boyd2004} is adopted to solve the optimization problem. 
The proposed inverse filtering algorithm is advantageous in that the sparsity promotion is efficient for noise reduction, and the tolerance for the $\ell_2$ fitting error can be automatically set according to the noise power, 
while the regularization term in \cite{hikichi2007,kodrasi2016,kodrasi2016robust} has to be empirically set even if the noise power is known. 
Finally, the time-domain signal is obtained by inverse STFT.

The remainder of this paper is organized as follows. The blind channel identification in the STFT-domain is given in Section~\ref{sec:ci}.
The inverse filtering method is presented for dereverberation and noise reduction in Section~\ref{sec:dereverberation}.  
Experiments with binaural simulation data and with multichannel real recordings are presented in Section~\ref{sec:experiments1} and~\ref{sec:experiments2}, respectively. Section~\ref{sec:conclusion} concludes the work.

\section{STFT-domain Channel Identification}\label{sec:ci}
 
We consider a two channel system. In the time domain, the noise-free microphone signals $x(n)$ and $y(n)$ are 
\begin{align}\label{eq:xn}
 x(n)=s(n)\star a(n), \quad y(n)=s(n)\star b(n), 
\end{align}
where $\star $ denotes convolution, $s(n)$ is a non-stationary source signal, e.g., speech, and $a(n)$ and $b(n)$ are the RIRs.
The cross-relation method \cite{xu1995} is not practical for the time-domain RIR identification due to the near common zeros problem and the length of the filters. 
In this section, we propose a STFT-domain channel filter identification algorithm.

\subsection{Problem Formulation in the STFT Domain}
\label{subsec:bf}

The STFT representation of the signal $x(n)$ is 
\begin{align}\label{eq:stft}
 x_{p,k} = \sum_{n=-\infty}^{+\infty} x(n)\tilde{w}(n-pL)e^{-j\frac{2\pi}{N}k(n-pL)},
\end{align}
where $p=0,\dots,P-1$ and $k=0,\dots,N-1$ denote the frame index and the frequency index, respectively, $\tilde{w}(n)$ denotes an analysis window, and $N$ and  $L$ denote the frame (window) length, and the frame step, respectively.
Equation~(\ref{eq:stft}) gives the overlap-add view of STFT. In the filter bank interpretation, the analysis window is considered as the low-pass filter, and $L$ as the decimation factor.

The cross-band filter model consists in representing the STFT coefficient $x_{p,k}$ as a summation over multiple convolutions (between the STFT-domain source signal and filter) across frequency bins.
Mathmatically, the linear time invariant system~(\ref{eq:xn}) can be written in the STFT domain as 
\begin{align}\label{xpk2}
 x_{p,k} = \sum_{k'=0}^{N-1}\sum_{p'=-C}^{Q-1} s_{p-p',k'} \; a_{p',k,k'},
\end{align}
and note that there is an equivalent expression for $y_{p,k}$.  From \cite{avargel2007}, if $L<N$, then $a_{p',k,k'}$ is non-causal, with $C = \lceil N/L \rceil -1$ non-causal coefficients. 
The number of causal filter coefficients  $Q$ is related to the reverberation time. Let $w(n)$ denote the STFT synthesis window.
The STFT-domain impulse response $a_{p',k,k'}$ is related to the time-domain impulse response $a(n)$ by:
\begin{align}\label{eq:hp}
a_{p',k,k'}={(a(n)\star \zeta_{k,k'}(n))}|_{n=p'L},
\end{align}
which represents the convolution with respect to the time index $n$ evaluated at frame steps, with
\begin{align}\label{phik}
\zeta_{k,k'}(n) = e^{j\frac{2\pi}{N}k'n}\sum_{m=-\infty}^{+\infty} \tilde{w}(m) \: w(n+m) \: e^{-j\frac{2\pi}{N}m(k-k')}.
\end{align}
To simplify the analysis, we consider the convolutive transfer function (CTF) approximation, i.e., only band-to-band filters with $k=k'$ are considered: 
\begin{equation}
\label{eq:xpk3}
 x_{p,k} \approx \sum\nolimits_{p'=-C}^{Q-1} s_{p-p',k}a_{p',k}= s_{p,k}\star a_{p,k},
 \end{equation}
and similarly $y_{p,k} \approx  s_{p,k}\star b_{p,k}$.   

To identify the filters $a_{p,k}$ and $b_{p,k}$, the cross-relation between the two channels
\begin{align}\label{eq:xyha}
 x_{p,k}\star b_{p,k}=s_{p,k}\star a_{p,k}\star b_{p,k}=y_{p,k}\star a_{p,k}
\end{align}
 can be used. This relation was originally proposed for the time-domain filter identification and here is extended to the CTF domain. The conditions that this identification problem has a unique solution are given in
\cite{xu1995}, namely that (i) the linear complexity of the source signal $s_{p,k}$ should be sufficiently large to fully excite the filters, and that (ii) the two filters $a_{p,k}$ and $b_{p,k}$ are co-prime, i.e. they do not share any common zeros.
The first condition can be satisfied by increasing the length of source signal $s(n)$. The second condition is related to the configuration of the STFT. 
Prior to the detailed filter identification algorithm, below we analyze the influence of the STFT configuration on the signal reconstruction, CTF approximation and the common zeros issue. 

\subsection{Analysis of STFT Configuration}\label{ssec:stft}

\begin{figure}[t]
\centering
{\includegraphics[width=0.8\columnwidth]{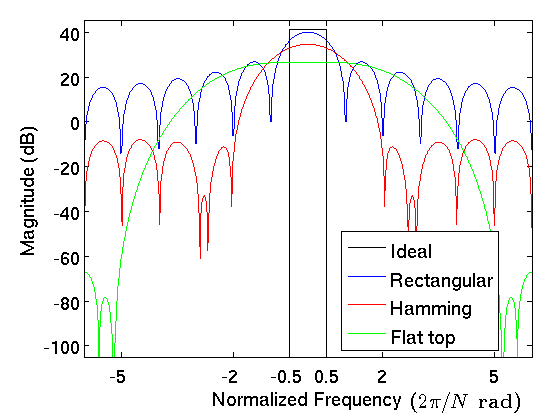}}
\caption{\small{The frequency response of STFT windows. }} 
\label{fig:window}
\vspace{-3mm}
\end{figure}

The filter bank interpretation of the STFT is that the time domain signal is first modulated (frequency shifted) by $e^{-j\frac{2\pi}{N}kn}$, then low-pass filtered with the analysis window $\tilde{w}(n)$, and down-sampled by $L$.
Let $\tilde{W}(\omega)$ denote the frequency responses of $\tilde{w}(n)$, where $\omega$ denotes the angular frequency. Fig.~\ref{fig:window} shows the frequency responses of three typical windows, i.e. rectangular, Hamming and flat-top windows. 
In addition, the ideal low-pass filter that has the bandwidth of $2\pi/N$ is shown with a black rectangle. 
The down sampling operation will fold the frequency response with the cycle of $2\pi/L$. For the ideal filter, there is no frequency aliasing up to the critical down-sampling, i.e. $L=N$. 
While for the three practical windows, frequency aliasing appears when $L>1$. Although the frequency aliasing can be cancelled by the synthesis windows as long as the completeness condition is met, 
we still expect the aliasing to be as small as possible to guarantee a high robustness for spectral modifications. The dereverberation algorithm intends to recover the source signal $s_{p,k}$ from the microphone signals $x_{p,k}$ and $y_{p,k}$.
In other words, the microphone signals are modified to a large extent, which will disturb the aliasing cancellation. 
To this end, we should use the window with a high side-lobe attenuation, such as the Hamming or flat-top windows. 
In addition, the cutoff frequency of the lowpass filter should not be larger than the folding frequency $\pi/L$. Here we consider the cutoff frequency to be at the first zero-crossing, namely the main lobe.
Taking the Hamming window as an example, one can see from Fig.~\ref{fig:window} that the first zero-crossing is at about $4\pi/N$, as a result, the decimation factor should be set as $L\le N/4$.
We denote with $L_{max}$ the maximum decimation factor that guarantees a small frequency aliasing, and $L_{max}=N/4$ for the Hamming window. 

To illustrate the reliability of the CTF approximation, we analyze the significance of the cross-band filters for various windows. According to~(\ref{eq:hp}), the undecimated STFT-domain filter $\tilde{a}_{n,k,k'}$ is obtained 
by setting $L=1$. Applying the discrete-time Fourier transform to $\tilde{a}_{n,k,k'}$ with respect to the time index $n$, the frequency response is given as, \cite{avargel2007}:
\begin{align}\label{eq:Ak}
 \tilde{A}_{k,k'}(\omega) = A(\omega)\tilde{W}\left(\omega-\frac{2\pi}{N}k\right)W\left(\omega-\frac{2\pi}{N}k'\right)
\end{align}
where $A(\omega)$ and $W(\omega)$ are the frequency responses of $a(n)$, $w(n)$, respectively. Without loss of generality, the synthesis window is assumed to possess the same frequency response as 
the analysis window. The product of $\tilde{W}(\omega-\frac{2\pi}{N}k)$ and $W(\omega-\frac{2\pi}{N}k')$ indicates the power of the cross-band filters. 
For the ideal filter, the cross band filters $\tilde{A}_{k,k'}(\omega)$ with $k'\ne k$ are equal to zero. In other words, the band-to-band filter $\tilde{A}_{k,k}(\omega)$ is a perfect STFT representation of the filter $a(n)$.
However, for the three practical windows, it can be seen from Fig.~\ref{fig:window} that $\tilde{A}_{k,k'}(\omega)$ are not zero for $k'\ne k$. As a result, the band-to-band filter, i.e. CTF, can only approximately represent the filter $a(n)$ in the STFT domain.  
The approximation error depends on the power of the cross-bands filters with $k'\ne k$ relative to the power of the band-to-band filter. 
Consider a specific frequency band $\omega=0$, it can be seen from Fig.~\ref{fig:window} that the rectangular window has the least approximation error, since the overlap between $\tilde{W}(0)$ and $W(\frac{2\pi}{N}k')|_{k'\ne 0}$ is the smallest among the three windows.  
The flat-top window has the largest approximation error.

The decimation will fold the frequency response, as for the band-to-band filter $a_{p,k}$ with the decimation factor $L$, the frequency response is
\begin{align}\label{eq:Akd}
 \tilde{A}_{k,k}(\omega)_{\downarrow L} = \frac{1}{L}\sum_{l=0}^{L-1} \tilde{A}_{k,k}\left(\frac{\omega}{L}-\frac{2\pi}{L}l\right), 
 \end{align}
 where $\tilde{A}_{k,k}(\omega)$ is defined in~(\ref{eq:Ak}) with $k'=k$. 
 To simplify the analysis, the magnitude of the side lobes is assumed to be zero. The decimation factor is set to be less than the maximum decimation factor $L_{max}$ to guarantee a small frequency aliasing. 
 Without loss of generality, we consider the case that $N/L$ is an integer, and the frequency band $\bar{k}$ being the integer multiple of $N/L$. Then~(\ref{eq:Akd}) is simplified as
\begin{align}\label{eq:Abkd}
 \tilde{A}_{\bar{k},\bar{k}}(\omega)_{\downarrow L} = \frac{1}{L}A\left(\frac{\omega}{L}-\frac{2\pi}{N}\bar{k}\right)\tilde{W}\left(\frac{\omega}{L}\right)W\left(\frac{\omega}{L}\right).
\end{align}
The filter $\tilde{W}(\frac{\omega}{L})$ involves only the mainlobe if $L=L_{max}$, and involves some sidelobes if $L<L_{max}$. 
For the ideal filter, the mainlobe is fully excited, and thence the filter $\tilde{W}(\frac{\omega}{L})$ is fully excited when $L=L_{max}=N$. 
However, for the three practical windows, the magnitude of $\tilde{W}(\frac{\omega}{L})$ is dramatically decreasing along with the increase of frequency, even in the mainlobe. 
Thence, the magnitudes of $\tilde{W}(\frac{\omega}{L})W(\frac{\omega}{L})$ are close to zero in the high frequency region. Note that for the case that $N/L$ is not an integer, 
and/or the frequency band being not the integer multiple of $N/L$, the zero-magnitude region also exsits possibly in a different frequency region. 
This zero-magnitude region is caused by the filter banks, and thence is also present in the frequency response of filter $b_{p,k}$. This common zeros of two channels are problematic for the cross-relation method~(\ref{eq:xyha}).
It can be seen from Fig.~\ref{fig:window} that the flat-top window has the least common zeros among the three windows, since its mainlobe is excited the most.

To summarize, extending the time-domain cross-relation method to the STFT domain is not a trivial task. It suffers from the problems of frequency aliasing, CTF approximation and common zeros. 
The windows (low-pass filters) and the frame step (decimation factor) are crucial for circumventing these problems. 
Briefly, the frequency aliasing can be suppressed by using a window with a high side lobe attenuation, and a small frame step. A small CTF approximation error requires the window to have a narrow main lobe. 
To avoid the common zeros, the window is better to have a rectangular main lobe. Otherwise, a larger frame step (even critical sampling) is needed to fully excite the CTF filter.
Unfortunately, these requirements can only be fulfilled by the ideal filter with critical sampling. For the practical filters, as discussed above, one specific STFT configuration can only fit parts of the conditions.

In this work, to achieve a good compromise between these requirements, we propose a novel STFT configuration. The Hamming window with frame step $L=N/4$ is adopt for the STFT of the signals, which has the negligible frequency aliasing and
acceptable CTF approximation error (slightly larger than the approximation error of the rectangular window). In addition, to avoid the common zeros problem, the STFT-domain filters $a_{p,k}$ and $b_{p,k}$ are forced to take the critical sampling, i.e. we set $L_f=N$, where $L_f$ denotes the frame step of the STFT-domain filters. 
The details will be presented in the next subsection. As for the window length, a relatively large one could be selected, e.g. 64~ms in this work. A large window size leads to a small number of channel filter taps, which is beneficial to channel identification. 
In addition, the spectral differences between the neighboring frequency bands get smaller with the increasing of the frequency resolution (window length), which will mitigate the cross-bands effect.

\subsection{Channel Identification}\label{ssec:si}

Note that since the channel identification algorithm is applied frequency-wise, hereafter the frequency index $k$ is omitted unless necessary.
In~(\ref{eq:xpk3}) and~(\ref{eq:xyha}), the frame index $p$ ($p\in[0,P-1]$ for the signals $s_{p}$, $x_{p}$ and $y_{p}$, and $p\in[-C,Q]$ for the filters $a_{p}$ and $b_{p}$) corresponds to the frame step $L=N/4$. 
Denote the filters in vector form as  $\mathbf{a}$ and $\mathbf{b}$. 
To avoid the common zeros problem, we further down sample the filters by a factor 4. The down sampled filters $a_{p\downarrow 4}$ and $b_{p\downarrow 4}$ correspond to the critical sampling, which have larger frequency aliasing than the original filters. 
There will be no non-causal coefficients anymore for the case of critical sampling. The down sampled filters start with the zeroth tap, and have a length of $\tilde{Q} = \lceil Q/4 \rceil$.  
They are written in vector form as 
\begin{align}\label{eq:abv}
 \tilde{\mathbf{a}} = [a_0,a_4,\dots,a_{4(\tilde{Q} -1)}]^{\top}, \nonumber \\
 \tilde{\mathbf{b}} = [b_0,b_4,\dots,b_{4(\tilde{Q} -1)}]^{\top}.
\end{align}
where $^{\top}$ is the transpose operator.  

Define the matrix $\mathbf{X}$ from the signal $x_{p}$ as
\begin{equation}
\mathbf{X} = 
\footnotesize{\begin{bmatrix}
x_0 & 0 & \cdots & \cdots & 0 \\
\vdots & \ddots & \ddots & \ddots & \vdots \\
x_{p} & x_{p-4} & \ddots & \ddots & x_{p-4(\tilde{Q}-1)} \\
\vdots & \ddots & \ddots & \ddots & \vdots \\
x_{P-1} & x_{P-5} & \cdots & \cdots & x_{p-1-4(\tilde{Q}-1)} \\
\end{bmatrix}}   
\end{equation}
with the size of $P\times\tilde{Q}$. The row number is the frame number of the oversampled signal, while the column number is the length of the critical sampled filter. 
This indicates that $\mathbf{X}\tilde{\mathbf{b}}$ is the convolution between the  oversampled signal and the critical sampled filter.
The matrix $\mathbf{Y}$ is defined from $y_{p}$ following the same principle. 
Then~(\ref{eq:xyha}) can be rewritten as $\mathbf{X}\tilde{\mathbf{b}}=\mathbf{Y}\tilde{\mathbf{a}}$, which is equivalent to
\begin{align}\label{eq:ep}
 \mathbf{Z}\mathbf{c}=0
\end{align}
where $\mathbf{Z}=[\mathbf{Y},-\mathbf{X}]$ and $\tilde{\mathbf{c}}=[\tilde{\mathbf{a}}^{\top},\tilde{\mathbf{b}}^{\top}]^{\top}$.  
In \cite{xu1995,huang2003,lin2012}, the filter vector $\tilde{\mathbf{c}}$ is estimated by taking an eigenvector of $\mathbf{Z}$ corresponding to a zero eigenvalue.
This method is available only for the case that the filter length is exactly known.  
However, in practice, the length of the RIR cannot be exactly measured.  
In addition, the one-dimensional null space of $\mathbf{Z}$ could be easily contaminated even by a mild noise interference. 
As a result, this eigenvector method is hardly applicable for RIR identification.

Instead of the  eigenvector method, we estimate the filter vector $\tilde{\mathbf{c}}$ by solving the following least-square problem
\begin{align}\label{eq:ls}
 \text{min}   \parallel \mathbf{Z}\tilde{\mathbf{c}} \parallel^2, \quad \text{s.t. }   \mathbf{g}^{\top}\tilde{\mathbf{c}}=1, 
\end{align}
where $\mathbf{g}$ is a constant vector to constrain $\tilde{\mathbf{c}}$. Empirically, $\mathbf{g}$ is set as
\begin{align}\label{eq:g}
\mathbf{g} = [1, \underbrace{0, \dots,0}_{\tilde{Q}-1}, 1, \underbrace{0, \dots,0}_{\tilde{Q}-1}]^{\top}.
\end{align}
Here we constrain the sum of the first entries of the two filters to 1. The constraint can be written as $\mathbf{g}^{\top}\tilde{\mathbf{c}}=a_0+b_0=1$. 
Constraining the scale of the first entries will remove the delay ambiguity, namely the direct-path responses is enforced to start with the first entries. Moveover, the first entries of the RIRs usually have a larger amplitude than the others in $\tilde{\mathbf{c}}$,
and the minimization of the objective function tends to suppress the scale of the unconstrained entries. Thence, constraining the scale of the first entries may promote a reasonable estimation of $\tilde{\mathbf{c}}$.
The solution of~(\ref{eq:ls}) is
\begin{align}\label{eq:solution}
\hat{\mathbf{c}}=\frac{(\mathbf{Z}^H\mathbf{Z})^{-1}\mathbf{g}}{\mathbf{g}^{\top}(\mathbf{Z}^H\mathbf{Z})^{-1}\mathbf{g}}
\end{align}
where $^H$ denotes complex transpose. Consequently, the estimations of $\tilde{\mathbf{a}}$ and $\tilde{\mathbf{b}}$ are respectively $\hat{\mathbf{a}}=\hat{\mathbf{c}}_{1:\tilde{Q}}$ and $\hat{\mathbf{b}}=\hat{\mathbf{c}}_{\tilde{Q}+1:2\tilde{Q}}$.
This channel filter identification method can be extended to the multichannel case. 
The filter vector $\tilde{\mathbf{c}}$ would then be the concatenation of all the channels, and the signal matrix $\mathbf{Z}$ can be organized as proposed in \cite{xu1995,gannot2003} that concatenates the signal matrices of each microphone pair. 
Correspondingly, in the constraint vector $\mathbf{g}$, we set the entries corresponding to the first entries of each channel to 1, and the others to 0. The solution is still (\ref{eq:solution}).


It is obvious that $\hat{\mathbf{a}}$ and $\hat{\mathbf{b}}$ are the estimations of a normalized version of $\tilde{\mathbf{a}}$ and $\tilde{\mathbf{b}}$, and the normalization factor is $a_0+b_0$. 
However, the normalization factor varies along the frequency bands, which leads to the gain ambiguity. 
To remove the gain ambiguity we propose to further normalize both the filters by the first entry of one of the filters, e.g. $\hat{a}_0$. 
Formally, the normalized filters are defined as $\bar{\mathbf{a}}=\hat{\mathbf{a}}/\hat{a}_0$ and $\bar{\mathbf{b}}=\hat{\mathbf{b}}/\hat{a}_0$, where $\bar{a}_0=1$. Obviously, $\bar{\mathbf{a}}$ and $\bar{\mathbf{b}}$ are the estimations of $\tilde{\mathbf{a}}$ and $\tilde{\mathbf{b}}$ normalized by $a_0$.
In the frequency band $k$, the source signal corresponding to $\tilde{\mathbf{a}}$ and $\tilde{\mathbf{b}}$ is $s_{p,k}$, thence the one corresponding to $\bar{\mathbf{a}}$ and $\bar{\mathbf{b}}$ is $a_{0,k}s_{p,k}$. 
From~(\ref{eq:hp}), $a_{0,k}$ can be represented as
\begin{align}
 a_{0,k} =\sum_{n=0}^{N-1} a(n)\nu(n)e^{-j\frac{2\pi}{N}kn},
\end{align}
where $\nu(n)=\sum_m\tilde{w}(m)w(m-n)$ is a window function. 
Therefore, $a_{0,k}|_{k=0}^{N-1}$ can be interpreted as the Fourier transform of the impulse response segment $a(n)|_{n=0}^{N-1}$ windowed by $\nu(n)$.
As a result, the time-domain signal corresponding to $a_{0,k}s_{p,k}$ will be the convolution between $s(n)$ and $a(n)|_{n=0}^{N-1}$. The gain ambiguity is removed by consistently normalizing the filters.

\section{Inverse Filtering for dereverberation}\label{sec:dereverberation}

In this section we propose to estimate the source signals in the STFT domain by inverse filtering. In addition, to exploiting the spectral sparsity of the source signal in the STFT-domain, 
the $\ell_1$-norm minimization is proposed to suppress the noise caused by the filter perturbations and the ambient noise interference. 

\subsection{Inverse Filtering}\label{ssec:if}

Let us still consider the two channel case. The filter $\bar{\mathbf{a}}$ is an estimation of the critical sampled and frequency aliased filter $\tilde{\mathbf{a}}$, from which we could reconstruct an estimation of the original filter ${\mathbf{a}}$ by interpolation. 
The \emph{first-order hold} interpolation,\footnote{Some other interpolation methods have also been tested, such as \emph{zero-order hold} interpolation, piecewise cubic spline interpolation. They achieve almost the same results. } namely linear interpolation, is applied. 
Note that the perfect reconstruction cannot be achieved due to the frequency aliasing of $\bar{\mathbf{a}}$. 
From the interpolated filter ${\mathbf{a}}$, we construct the filter matrix $\mathbf{A}$ as 
\begin{equation}
\mathbf{A} = 
\footnotesize{\begin{bmatrix}
a_0 & 0     &  \cdots & \cdots& \cdots  &\cdots & 0  \\
a_1 & a_0 & \ddots& \ddots  &\ddots & \ddots & 0  \\
\vdots & \ddots & \ddots& \ddots & \ddots & \ddots & \vdots \\
a_{Q-1} & a_{Q-2} & \ddots & {a}_0 & 0 & \ddots  & 0 \\
0 & a_{Q-1} & a_{Q-2} & \ddots & {a}_0 & \ddots & 0 \\
\vdots & \ddots & \ddots & \ddots & \ddots & \ddots  & \vdots \\
0& \cdots &0 & a_{Q-1} & a_{Q-2} & \cdots & a_0 \\
\end{bmatrix}}  \nonumber
\end{equation}
of size of $P\times P$. In $\mathbf{A}$, the transpose of the filter ${\mathbf{a}}$ is duplicated as the row vectors, with one element shift per row.  
The filter matrix $\mathbf{B}$ is defined from ${\mathbf{b}}$ following the same principle. Then we concatenate the two matrices to yield $\mathbf{C}=[\mathbf{A}^{\top},\mathbf{B}^{\top}]^{\top}$.

We define the microphone signals $x_p$ and $y_p$ in vector form as $\mathbf{x}=[x_0,\cdots,x_{P-1}]^{\top}$ and $\mathbf{y}=[y_0,\cdots,y_{P-1}]^{\top}$, respectively. 
In the previous section we assumed that the microphone signals were noise free. 
In this section, we explicitly introduce additive noise to the microphone signals as $\mathbf{x}=\mathbf{x}_c+\mathbf{e}_x$, 
where $\mathbf{x}_c$ and $\mathbf{e}_x$ denote the noise-free signal and noise, respectively. The noise-free signal and noise are assumed to be uncorrelated. The noise signal obeys an i.i.d. complex Gaussian distribution.
Similarly, define $\mathbf{y}=\mathbf{y}_c+\mathbf{e}_y$.  Then we concatenate them as $\mathbf{z}_c=[\mathbf{x}_c^{\top},\mathbf{y}_c^{\top}]^{\top}$, $\mathbf{e}=[\mathbf{e}_x^{\top},\mathbf{e}_y^{\top}]^{\top}$ and 
$\mathbf{z}=[\mathbf{x}^{\top},\mathbf{y}^{\top}]^{\top}=\mathbf{z}_c+\mathbf{e}$. Note that the proposed channel identification algorithm was directly applied to the noisy case.

We define the source signal $s_p$ in vector form as $\mathbf{s}=[s_0,\cdots,s_{P-1}]^{\top}$.  The complex source signal can be estimated by minimizing the $\ell_2$-norm $\parallel \mathbf{C}\mathbf{s} -\mathbf{z}\parallel$.
Due to the CTF approximation error and the frequency aliasing of the critical sampled filters, the filter estimations are inaccurate.
We found that the amplitude of the filters is reliable, while the phase information is very inaccurate. Therefore, we only use the amplitude of the filters to recover the amplitude of the source signal.
We define $\tilde{\mathbf{z}}=|\mathbf{z}|$ and $\tilde{\mathbf{C}}=|\mathbf{C}|$, where $|\cdot|$ denotes the entry-wise absolute value of a matrix or vector.
The amplitude filtering is defined as $\tilde{\mathbf{C}}\mathbf{s}|_{\mathbf{s}\succeq \mathbf{0}}$, where $\succeq$ denotes entry-wise vector inequality and $\mathbf{0}$ is the $P$-dimensional vector with all entries equal to 0. 
The amplitude of the source signal can be recovered by minimizing the following cost function
\begin{align}\label{eq:l2} 
 \bar{\mathbf{s}} = \mathop{\textrm{argmin}}_{\mathbf{s}, \ \text{s.t.} \ \mathbf{s}\succeq \mathbf{0} } \parallel \tilde{\mathbf{C}}\mathbf{s} -\tilde{\mathbf{z}}\parallel^2. 
\end{align}
For the ideal noise-free case with the well-estimated filter $\mathbf{C}$ and the true source signal $\mathbf{s}$, the relation $\mathbf{C}\mathbf{s}\approx\mathbf{z}$ holds. Based on the triangle inequality theorem, we have $|\mathbf{C}||\mathbf{s}|\ge|\mathbf{z}|$.
Thence, the solution of (\ref{eq:l2}) would be an underestimation of $|\mathbf{s}|$. 
The underestimated source singal is smaller than the proper estimation, thus it will possibly suffer from some distortions, but it will not include more reverberations than the proper estimation.
In addition, as mentioned in Section \ref{ssec:si}, corresponding to the normalized filters, the source signal involves some early reflections, which provides more tolerance to the underestimation. 
Briefly stated, the lost information of the direct-path signal due to the underestimation could possibly be preserved in its early reflections, and vice versa. 
The experiments using both the estimated CTFs and the true CTFs computed by (\ref{eq:hp}) demonstrate that the distortion of source signal is very low (for details see Section~\ref{ssec:kctf}).

In addition, the STFT-domain signal is sparse, which can be exploited by minimizing the $\ell_1$ norm of the source signal to suppress the noise. For the positive amplitude vector $\mathbf{s}$, the $\ell_1$ norm is $|\mathbf{s}|_1=\mathbf{1}^{\top}\mathbf{s}, \mathbf{s}\succeq \mathbf{0}$, 
where $\mathbf{1}$ is the $P$-dimensional vector with all entries set to 1. The $\ell_1$ minimization can be realized by adding the $\ell_1$ regularization on~(\ref{eq:l2}), however the regularization factor is difficult to be automatically set.
We realize the $\ell_1$ minimization by solving the constrained optimization problem:
\begin{align}\label{eq:l1} 
 \tilde{\mathbf{s}} &= \mathop{\textrm{argmin}}_{\mathbf{s}} \mathbf{1}^{\top}\mathbf{s}  \\
 \text{s.t.} & \quad  \mathbf{s}\succeq \mathbf{0}, \ \parallel \tilde{\mathbf{C}}\mathbf{s} -\tilde{\mathbf{z}}\parallel^2\le \delta. \nonumber
\end{align}
The $\ell_2$ fit $\parallel \tilde{\mathbf{C}}\mathbf{s} -\tilde{\mathbf{z}}\parallel^2$ is relaxed to at most the tolerance $\delta$. 

The relaxing tolerance $\delta$ is related to the noise power in the microphone signal. 
Let $\sigma_{ex}^2$ and $\sigma_{ey}^2$ denote the noise PSD in the two channels, which can be estimated from the pure noise signal for stationary noise, or estimated by a noise PSD estimator for non-stationary noise, e.g. \cite{mine2016assp,gerkmann2012}.  
The squared $\ell_2$ norm of the noise signal $\parallel \mathbf{e}_x \parallel^2$ (or $\parallel \mathbf{e}_y \parallel^2$) follows an Erlang distribution with mean $P\sigma_{ex}^2$ (or $P\sigma_{ey}^2$) and variance $P\sigma_{ex}^4$ (or $P\sigma_{ey}^4$). 
Assume the noise signal in the two channels are uncorrelated, then $\parallel \mathbf{e}\parallel^2$ has the mean $P(\sigma_{ex}^2+\sigma_{ey}^2)$ and variance $P(\sigma_{ex}^4+\sigma_{ey}^4)$. 
To relax the $\ell_2$ fit to the noise power, we set the noise relaxing term as 
\begin{align}
 \delta_e=P(\sigma_{ex}^2+\sigma_{ey}^2)-2\sqrt{P(\sigma_{ex}^4+\sigma_{ey}^4)}.
\end{align} 
Subtracting two times the standard deviation makes the probability that the $\ell_2$ fit being larger than $\parallel \mathbf{e}\parallel^2$  very small. 
When the $\ell_2$ fit  is allowed to be larger than $\parallel \mathbf{e}\parallel^2$, the minimization of $\mathbf{1}^{\top}\mathbf{s}$ will distort the source signal. Here we prefer less source signal distortion on more noise reduction.  
As a result, some noise remains in the estimated source signal. 

\setlength{\tabcolsep}{3pt}
\renewcommand\arraystretch{1.3}
\begin{table*}[t]
\centering\small
\caption{{The specifications of the optimization problems (\ref{eq:l2}) and (\ref{eq:l1}), where $\mathbf{I}\in\mathbb{R}^{P\times P}$ denotes the identity matrix, $\mathbf{0}_{P\times P}\in\mathbb{R}^{P\times P}$ 
denotes the matrix with all entries set to 0.}}
\label{tab:spe}
\begin{tabular}{|c|c| c | c | c|c|c|c| }
\hline       
&$f_0(\mathbf{s})$	    &$f_i(\mathbf{s})$          & $\mathbf{\lambda}$       & $\nabla f_0(\mathbf{s})$ &   $\nabla^2 f_0(\mathbf{s})$   & $Df(\mathbf{s})$  &$\nabla^2 f_i(\mathbf{s})$         \\  \hline 
(\ref{eq:l2})&$\parallel \tilde{\mathbf{C}}\mathbf{s} -\tilde{\mathbf{z}}\parallel^2$	    &$-s_i, i\in[1,P]$    &$\lambda_i,i\in[1,P]$   & $\tilde{\mathbf{C}}^{\top}(\tilde{\mathbf{C}}\mathbf{s} -\tilde{\mathbf{z}})$ & $\tilde{\mathbf{C}}^{\top}\tilde{\mathbf{C}}$ & $-\mathbf{I}$ & $\mathbf{0}_{P\times P},\forall i$ \\ \hline   
(\ref{eq:l1})&$\mathbf{1}^{\top}\mathbf{s}$	    &$-s_i, i\in[1,P]$;                    & $\lambda_i,i\in[1,P];$   &$\mathbf{1}$  & $\mathbf{0}_{P\times P}$ &  $[-\mathbf{I},\tilde{\mathbf{C}}^{\top}(\tilde{\mathbf{C}}\mathbf{s} -\tilde{\mathbf{z}})]^{\top} $  & $\mathbf{0}_{P\times P}, i\in[1,P]$;    \\   
 &    &$\parallel\tilde{\mathbf{C}}\mathbf{s} -\tilde{\mathbf{z}}\parallel^2-\delta, i=P+1$ &$\lambda_{P+1}$ & & & & $\tilde{\mathbf{C}}^{\top}\tilde{\mathbf{C}}$,  $i=P+1$ \\ \hline
\end{tabular}
\vspace{-.3cm}
\end{table*}
\renewcommand\arraystretch{1.0}

Besides, the $\ell_2$ fit should also be relaxed with respect to the filter perturbations, since the amplitude filtering $\tilde{\mathbf{C}}\mathbf{s}$ is not accurate to fit $|{\mathbf{z}}_c|$ by definition. 
The inaccuracy is akin to the level of the noise-free signal, i.e. $\lambda_c=\parallel \mathbf{z}_c\parallel^2$. It can be estimated by spectral subtraction as 
\begin{align}
 \hat{\lambda}_c=\max(\parallel \mathbf{z}\parallel^2-P(\sigma_{ex}^2+\sigma_{ey}^2), \ 0).
\end{align} 
Empirically, the relaxing term with respect to the noise signal is set to $\delta_c=0.05\hat{\lambda}_c$.  

The relaxing tolerance can be set by $\delta_e+\delta_c$.
However, for this quantity, the $\ell_2$ norm constraint in~(\ref{eq:l1}) is not definitely feasible, since both $\delta_e$ and $\delta_c$ are set to be relatively small to avoid the source signal distortion. 
The minimum $\ell_2$ norm fitting error is defined in~(\ref{eq:l2}), i.e. $\parallel \tilde{\mathbf{C}}\bar{\mathbf{s}} -\tilde{\mathbf{z}}\parallel^2$, which is taken as the lower bound of the relaxing tolerance.
Overall, the relaxing tolerance is set as
\begin{align}
 \delta=\max(\delta_e+\delta_c,\ 1.05\parallel \tilde{\mathbf{C}}\bar{\mathbf{s}} -\tilde{\mathbf{z}}\parallel^2)
\end{align} 
where $1.05$ is a slack factor. 

The extension of this inverse filtering model to the multichannel case is straightforward. The filter matrix $\mathbf{C}$ and signal vector $\mathbf{z}$ are constructed by concatenating the filter matrices and the microphone signal vectors of all the channels, respectively.

The optimal solution $\tilde{\mathbf{s}}$ is an estimation of the amplitude of the source signal. The elements in $\tilde{\mathbf{s}}$ are written with the frame and the frequency index as $\tilde{s}_{p,k}$.
The phase of one of the microphone signals is taken as the phase of the estimated source signal, we have $\hat{s}_{p,k}=\tilde{s}_{p,k}e^{j\arg[x_{p,k}]}$, where $\arg[\cdot]$ is the phase of complex number.
 The time-domain source signal $\hat{s}(n)$ can be obtained Apply the inverse STFT to $\hat{s}_{p,k}$.
As mentioned in Sec. \ref{ssec:si}, based on the $a_{0,k}$-normalized filters, the output of the inverse filtering $\hat{s}_{p,k}$ is an estimation of $a_{0,k}s_{p,k}$. 
Consequently, the time-domain signal $\hat{s}(n)$ is an estimation of $s(n)\star a(n)|_{n=0}^{N-1}$, where $a(n)$ starts with the direct-path impulse response. 
The window size $N$ is generally far larger than the duration of the direct-path impulse response, thus $\hat{s}(n)$ involves the early reflections, e.g. 64 ms in this work.

\subsection{Convex Optimization}

Both (\ref{eq:l2}) and (\ref{eq:l1}) are convex optimization problems with an inequality constraint. We adopt the primal-dual interior-point method (PDIPM) \cite{boyd2004} to solve them. The book \cite{boyd2004} provides
a general optimization algorithm for a convex objective function $f_0(x)$ with a set of inequality constraints of the form $f_i(x) \le 0, i\in[1,m]$ and an affine equality constraint.
Here $x$ denotes the optimization variable and $m$ is the number of inequality constraints. Note that there is no affine equality constraint in the presented problems. Define the vector $f(x)=[f_1(x),\cdots,f_m(x)]^{\top}$ including all the inequality functions, 
and its derivative matrix $Df(x)=[\nabla f_1(x),\cdots,\nabla f_m(x)]^{\top}$, where $\nabla$ denotes gradient operator. Let $\lambda_i$ denote the dual variable corresponding to the inequality constraint $f_i(x) \le 0$.
The dual variable vector is $\lambda=[\lambda_1,\cdots,\lambda_m]$. In PDIPM, the inequality constraint is approximately formulated as an equality constraint by the logarithmic barrier function. The parameter $t$ sets 
the accuracy of the logarithmic barrier approximation, the larger $t$, the better the approximation.

The PDIPM is summarized in Algorithm~\ref{alg:PDIPM}, with variable update in Step 4 given by:
\begin{align}\label{eq:upd}
\begin{bmatrix}
\Delta x \\
\Delta \lambda \\
\end{bmatrix}
= - &
 \begin{bmatrix}
\nabla^2f_0(x)+\sum_{i}\lambda_i\nabla^2f_i(x) & Df(x)^{\top}  \\
-\text{diag}(\lambda)Df(x)^{\top} & -\text{diag}(f(x)) \\
\end{bmatrix}^{-1} \nonumber \\
&\times 
\begin{bmatrix}
\nabla f_0(x)+Df(x)^{\top} \\
-\text{diag}(\lambda)f(x)-(1/t)\mathbf{1}\\
\end{bmatrix}. 
\end{align}
In Algorithm~\ref{alg:PDIPM}, the so-called surrogate duality gap $\hat{\eta}^{(n)}$ is decreasing with the iterations, thence the parameter $t$ is increased by the factor $\mu$ (a positive value of the order of 10). 
The goal of the line search (Step 3) is to find the largest step-length $\zeta^{(n)}$ under the conditions that (i)~the updated dual variable $\lambda$ is nonnegative, (ii)~the inequality constraint is feasible with the updated variables, 
and (iii)~the problem is optimized. We refer to \cite{boyd2004} for more details. 
The convergence criterion is briefly set as (i) the surrogate duality gap $\hat{\eta}^{(n)}$ is less than a small tolerance $\epsilon$ to guarantee a high optimization, (ii) the dual residual $\nabla f_0(x)+Df(x)^{\top}$ is less than a small tolerance $\epsilon_{feas}$ to 
guarantee the feasibility of the variables. 

To apply the PDIPM to the problems (\ref{eq:l2}) and (\ref{eq:l1}), the general quantities in Algorithm~\ref{alg:PDIPM} should be accordingly specified. 
Table~\ref{tab:spe} gives the specifications for both (\ref{eq:l2}) and (\ref{eq:l1}).  
For solving (\ref{eq:l2}), a good initialization is to set $\mathbf{s}^{(0)}=|\mathbf{x}|$ and $\lambda^{(0)}$ is set to an arbitrary positive vector ($10\cdot\mathbf{1}_S$ in this paper).
For solving (\ref{eq:l1}), a feasible initialization is to set $\mathbf{s}^{(0)}$ and $\lambda^{(0)}$ as the solution of (\ref{eq:l2}).  

\begin{algorithm}
 \caption{\label{alg:PDIPM} Primal-dual interior-point method}
\begin{algorithmic} 
 \STATE Iteration step $n=0$. Initialize $f_i(x^{(0)})\le 0,\lambda_i^{(0)}\ge 0, \forall i$.
 \REPEAT 
 \STATE 1 Compute $\hat{\eta}^{(n)}=-f(x)^{\top}\lambda$,
 \STATE 2 Set $t^{(n)}:=\mu m/\hat{\eta}^{(n)}$,
 \STATE 3 Line search the step-length $\zeta^{(n)}$,
 \STATE 4 Update variables $x^{(n+1)}=x^{(n)}+\zeta^{(n)}\Delta x^{(n)}$, \\ \quad and $\lambda^{(n+1)}=\lambda^{(n)}+\zeta^{(n)}\Delta \lambda^{(n)}$.
 \UNTIL Convergence.
\end{algorithmic}
\end{algorithm}

\section{Experiments with Simulated Binaural Data} \label{sec:experiments1}


\setlength{\tabcolsep}{6pt}
\begin{table*}[t]
\centering
\caption{The performance measures for the noise-free microphone signals as a function of filter length. The filter length is set up to approximately $T_{60}$. The `unproc.' column is the corresponding performance measure of the microphone signal.
      The best performance measures for each condition are shown in \textbf{bold}.}
\label{tab:fl}

\begin{tabular}{c c | c | c  c c c c c c c c c c  }   
	    &                 &         & \multicolumn{11}{c}{Filter Length (ms)}  \\ \vspace{1mm}
	    & $T_{60}$ (s)    &unproc.  & 128  &    192  &    256  &    320  &    384  & 448  & 512  & 576  & 640  & 704  & 768  \\ \hline 
SRMR	    &0.5              &2.42     & 3.11 &    3.46 &    3.56 &    3.59 &    3.62 &    3.64  &   \textbf{3.66} & -&-&-&-     \\ \vspace{2mm}
	    &0.79             &1.93     & 2.60 &    3.05 &    3.21 &    3.28 &    3.32 &    3.35 &    3.37 &    3.38 &    3.39 & 3.40 &  \textbf{3.41}      \\  \hline  		    
PESQ	    &0.5              &2.44     & 2.60 &    2.70 &    \textbf{2.71} &    2.68 &    2.62 &    2.56 &    2.51 & -&-&-&-  \\   \vspace{2mm}
	    &0.79             &2.14     & 2.30 &    2.42 &    2.47 &    \textbf{2.49} &    2.48 &    2.45 &    2.40 &    2.36 &    2.32 & 2.28 &  2.24  \\   \hline  
LSD 	    &0.5              &3.95     & 3.02 &    \textbf{2.93} &    2.99 &    3.10 &    3.22 &    3.31 &    3.40 & -&-&-&-      \\ 
 	    &0.79             &5.15     & 4.03 &    3.63 &    3.48 &    \textbf{3.44} &    3.47 &    3.53 &    3.60 &    3.68 &    3.75 & 3.80 &  3.87      \\   
\end{tabular}
\end{table*}

In this section, we present a series of experiments with simulated binaural data. 
A set of BRIRs were generated with the ROOMSIM simulator \cite{campbell2004} combined with the head-related impulse responses (HRIRs) of the KEMAR dummy head \cite{gardner1995}.
The simulated room is of dimension $5$~m $\times$ $8$~m $\times$ $3$~m. The dummy head is located at ($1$~m, $4$~m, $1.5$~m). 
Sound sources are placed in front of the dummy head with azimuths (relative to the dummy head center) varying from $-90^\circ$ to $90^\circ$, spaced by $5^\circ$ (hence $37$ azimuths), and an elevation of $0^\circ$.
The dummy-head-to-source distances were always $2$~m. Two reverberation times, i.e. $T_{60} = 0.5$~s and $0.79$~s, are simulated by adjusting the absorption coefficients of the walls. 
The speech signals from the TIMIT dataset \cite{garofolo1988} are taken as the source signals, with the duration about 4~s. For each reverberation time, 50 TIMIT signals are convolved with the BRIRs with a random azimuth as the reverberated microphone signals.  
To generate the noisy microphone signal, the spatially uncorrelated stationary speech-like noise is added to the microphone signals with signal-to-noise ratio (SNR) of 0, 5, 10, 15, 20 dB, respectively. 

The sampling rate is 16 kHz. As already metioned in~\ref{ssec:stft}, the STFT takes the Hamming window, with the window length and frame step of $N=1,024$ (64~ms) and $L=N/4=256$, respectively.
In the convex optimization algorithm, the parameters are set to $\mu=20$, $\epsilon=\epsilon_{feas}=10^{-6}$. 
Since the early reflections are preserved in the proposed method, the reference (target) signal of the proposed method is set as the early (64 ms) reverberated signal, which is generated by convolving the source signal with the first 64 ms (starting from the direct-path) of the first-channel RIRs.
The noise PSD is estimated from the pure noise signals for various SNRs.

Two STFT-based state-of-the-art methods are compared, namely the weighted prediction error (WPE) \cite{yoshioka2011} method, and the coherent-to-diffuse power ration (CDR) method \cite{schwarz2015}. 
For the WPE method, under the conditions with $T_{60} = 0.5$~s and $0.79$~s, the number of filter coefficient is set to 50 and 80, respectively, which correspond  $0.8 T_{60}$ with respect to an 8~ms frame step.  
The source signal is taken as the reference signal. 
For the CDR method, the estimator with known DOA and unknown noise coherence is adopted, and the true DOA is used.
The early (50 ms) reverberated signal is taken as the reference signal as used in \cite{schwarz2015}.

Three metrics are taken to quantitatively evaluate the performance, (i) a non-intrusive metric, normalized speech-to-reverberation modulation energy ratio (SRMR) \cite{santos2014}, and two intrusive metrics (ii) 
perceptual evaluation of speech quality (PESQ) \cite{rix2001} and (iii) log-spectral distance (LSD) \cite{schwartz2016} with the desired dynamic range of 50 dB. 
The dereverberated signal of various methods is an estimation of their corresponding reference signal up to a time shift and/or a gain factor. 
Therefore, as is for PESQ, the signals are aligned and amplitude normalized prior to the computation of the LSD.  
For the SRMR and PESQ, the higher the better, and for LSD, the lower the better.

\subsection{Selection of Channel Filter Length}

The length of the channel filter $Q$ (or $\tilde{Q}$) corresponding to the frame step $L$ (or $L_f$) is a key parameter to be set. On the one hand, it influences the performance measures of the proposed method 
to a large extent. On the other hand, it is the only one prior knowledge that the proposed method requires, since it is related to the reverberation time.  

Table~\ref{tab:fl} shows the performance measures of the proposed method as a function of filter length.  
It is observed that SRMR increases along with the increasing of filter length. The reason is that SRMR mainly measures the amount of the reverberation, and a long filter leads to a more sparse source signal estimation.
However, when the channel filter is too long, the tail of the estimated filters will be noisy, which will distort the early reverberated signal. 
As a result, the PESQ score and LSD score degrade for the case of very long filter. 
When the filter length is set to be small, such as 128 ms, the filters can not cover most of the  energy of the real RIRs. Consequently, the fitler estimation is inaccurate, and all the three metrics performs poorly.  
As a good compromise of the different metrics, the filter length is set to approximately 0.5 times $T_{60}$, e.g. 256 ms for $T_{60}=0.5$ s and 384 ms for $T_{60}=0.79$ s. 
With this setting, the channel filters cover major part of the energy of the RIRs, and avoid the long tail effect.
Correspondingly, $Q$ ($\tilde{Q}$) is set to 16 and 24 (4 and 6) for the cases of $T_{60}=0.5$ s and $T_{60}=0.79$ s, respectively.
Note that although the performance scores are considerably affected by the selection of $Q$, the sound qualities by human listening are quite similar for the filter length selections from $0.3 T_{60}$ to $T_{60}$.

\begin{figure*}[t]
\centering
\subfloat[Source signal]{\includegraphics[width=0.32\textwidth,height=0.21\textwidth]{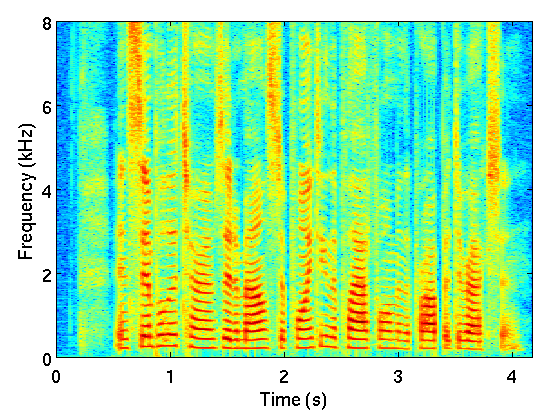}}
\subfloat[Early reverberated signal]{\includegraphics[width=0.32\textwidth,height=0.21\textwidth]{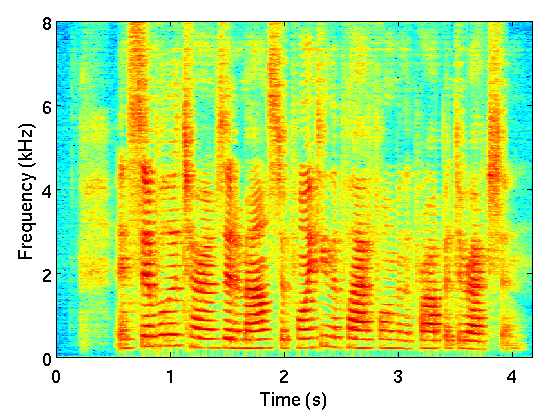}} 
\subfloat[Noisefree microphone signal]{\includegraphics[width=0.32\textwidth,height=0.21\textwidth]{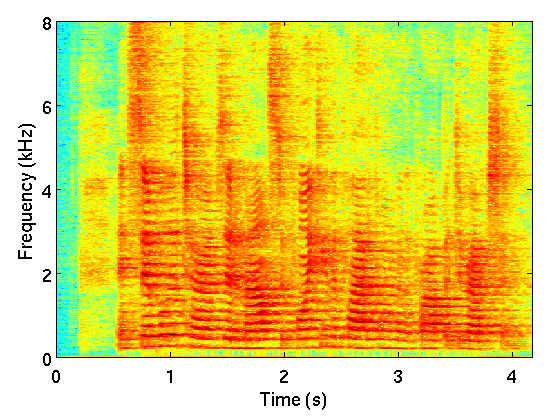}} \vspace{-0.3cm}\\ 
\subfloat[Output: true CTFs]{\includegraphics[width=0.32\textwidth,height=0.21\textwidth]{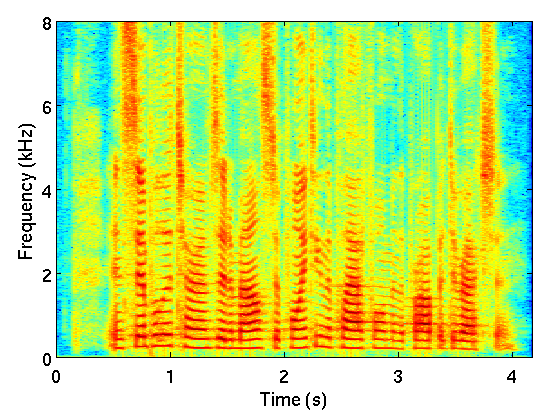}} 
\subfloat[Output: amplitude of true CTFs]{\includegraphics[width=0.32\textwidth,height=0.21\textwidth]{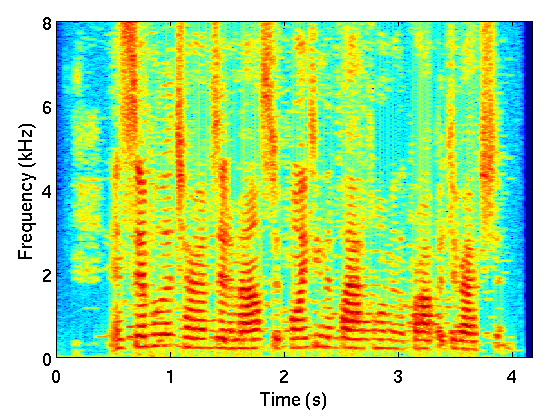}}
\subfloat[Output: amplitude of identified CTFs]{\includegraphics[width=0.32\textwidth,height=0.21\textwidth]{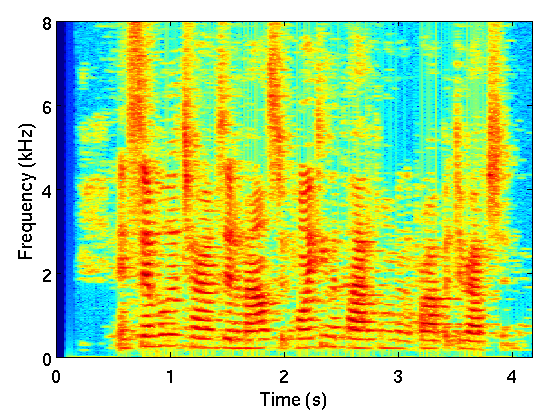}} 
\caption{Sonogram examples for the dereverberation algorithm using the true CTFs and identified CTFs. $T_{60}=0.79$~s.} 
\label{fig:knwonctf}
\vspace{-0.2cm}
\end{figure*}

\subsection{Comparison with the Knwon CTFs Case}\label{ssec:kctf}
This paper proposes an STFT domain blind channel identification method. As mentioned in the methodology part, the identified frequency-wise filter suffers from some errors i) the CTF approximation error, namely the loss of cross-band information, 
ii) the frequency aliasing of filters caused by the critical sampling, iii) the amplitude approximation error, namely the loss of phase information. To evaluate the applicability of the proposed method, 
we compare the dereverberated signals obtained by different approaches, i) using the true CTFs computed by~(\ref{eq:hp}) based on the known time domain filter, with the frame step $L=N/4$. 
The inverse filtering is carried out by minimizing the regualarized $\ell_2$-norm cost function  $\parallel \mathbf{C}\mathbf{s} -\mathbf{z}\parallel+\lambda|\mathbf{s}|$, 
where the regualarization term $|\mathbf{s}|$ is used to impose the sparsity of the source signal, and $\lambda$ is set to $10^{-3}$. The optimization method was proposed in \cite{mine2017assp} for the multi-source case, and is directly used for the single source case in this experiment.
This filter only has the CTF approximation error, but not the frequency aliasing and the loss of phase, ii) using the true CTFs and the inverse filtering method proposed in Sec. \ref{sec:dereverberation}, namely only the amplitude of the true CTFs is used, iii) the proposed blind dereverberation method. 

Fig.~\ref{fig:knwonctf} depicts sonogram examples for these three cases. Fig.~\ref{fig:knwonctf}a, b and c respectively illustrates the source signal, early reverberated signal and microphone singal. 
The smearing effect is evident by observing the microphone signal (Fig.~\ref{fig:knwonctf}c). 
From Fig.~\ref{fig:knwonctf}d, e and f, it can be seen that the signals of all the three approaches are much less reverberated than the microphone signal, showing the efficiency of the CTF-based inverse filtering method.  
The CTFs computed from the time domain filter do not have the gain ambiguity across frequencies. In other words, the frequency-wise dereverberated signals are consistent with the source signal, thus the output of true CTFs is an estimation of the source signal.
We can observe that the dereverberated signal of true CTFs (Fig.~\ref{fig:knwonctf}d) is close to the source signal, namely it does not include early reflections. 
This can also be verified by listening to the audio file, however, we can perceive a small delayed replica of the original source signal. 
This small replica is not obvious in the sonogram, and possibly caused by the loss of cross-band information.   
By only using the amplitude of true CTFs (Fig.~\ref{fig:knwonctf}e), it can be seen that it recovers the main structure of the source signal. 
There are some spectral distortions for the low power region due to the loss of phase information, however we could not clearly perceive these distortations by listening. 
The small replica still exists, but is perceively less natural than the output of true CTFs, possibly also due to the loss of phase information. 
Compared to Fig.~\ref{fig:knwonctf}d, it is observed that Fig.~\ref{fig:knwonctf}e has more speech distortions, but not more reverberations. This verifies the assertion that the underestimated source singal will only suffer from more distortions, but not from more reverberations. 

\begin{figure*}[t]
\centering
\begin{minipage}{.32\textwidth}
\subfloat[Output of WPE]{\includegraphics[width=1\textwidth,height=0.6562\textwidth]{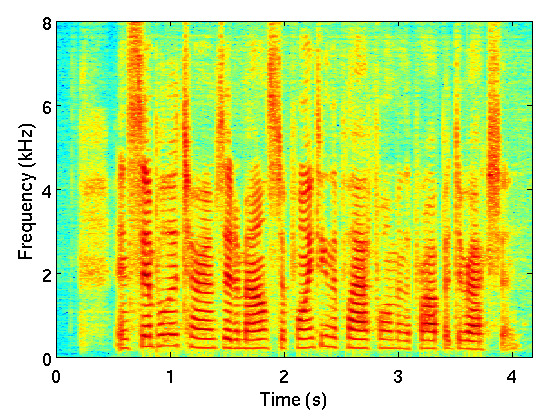}} \vspace{-.3cm} \\ 
\subfloat[Output of CDR]{\includegraphics[width=1\textwidth,height=0.6562\textwidth]{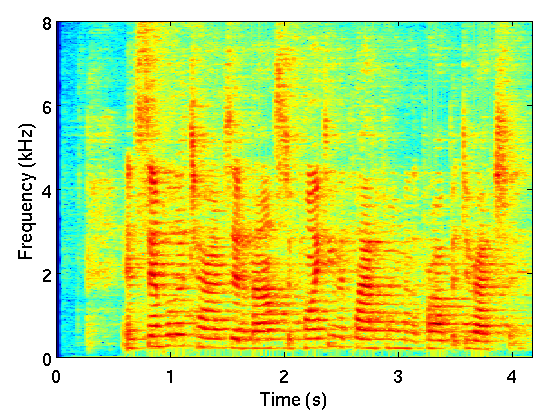}} 
\caption{Sonogram examples of WEP and CDR. $T_{60}=0.79$~s. The source signal and microphone signal are the ones in the example Fig.~\ref{fig:knwonctf}. } 
\label{fig:wpecdr}
\end{minipage}\hspace{0.6cm}
\begin{minipage}{.64\textwidth}
\vspace{-0.6 cm}
\subfloat[Noisy microphone signal]{\includegraphics[width=0.5\textwidth,height=0.3281\textwidth]{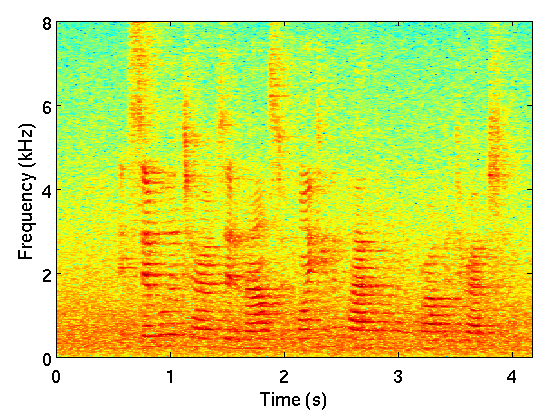}}
\subfloat[Output of WPE]{\includegraphics[width=0.5\textwidth,height=0.3281\textwidth]{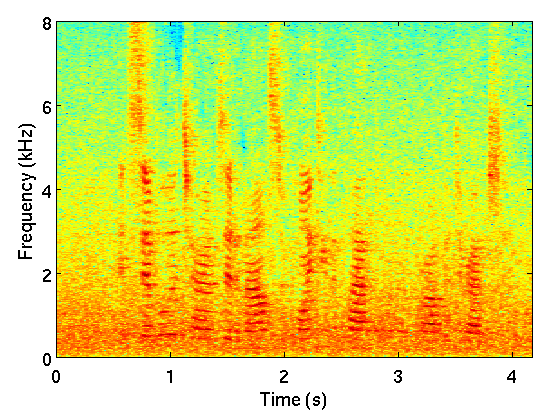}} \vspace{-.3cm} \\ 
\subfloat[Output of CDR]{\includegraphics[width=0.5\textwidth,height=0.3281\textwidth]{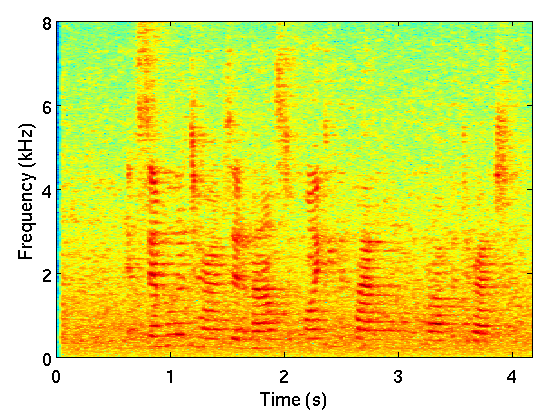}} 
\subfloat[Output of the proposed method]{\includegraphics[width=0.5\textwidth,height=0.3281\textwidth]{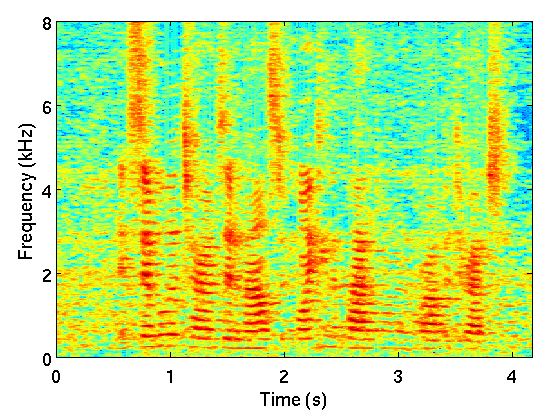}} 
\caption{Sonogram examples for the noisy case. The noisy microphone signal is generated by adding the speech-like noise to the noise-free microphone signal in Fig.~\ref{fig:knwonctf}c with SNR of 5 dB. } 
\label{fig:noisy}
\end{minipage}
\vspace{-0.2cm}
\end{figure*}


The output of the proposed blind method is shown in Fig.~\ref{fig:knwonctf}f, which recovers the main structure of the early reverberated signal (Fig.~\ref{fig:knwonctf}b). 
This verifies that the STFT configuration with oversampled signals and critical-sampled filters is suitable for the STFT-domain cross-relation method. 
This signal involves the early reflections, and also has some spectral distortions for the low power region. 
The sparsity regularization can partially reduce the residual noise caused by the filter aliasing and perturbations, and reduce the low power spectra in the replica. 
As a result, the overall residual noise is smaller, but more unnatural, and even sounds like musical noise. 
Compared to the singals in Fig.~\ref{fig:knwonctf}d and e, the output signal of the proposed blind method sounds less noisy, since the early reflections enhance the desired signal. 
Fig.~\ref{fig:knwonctf}f looks less distorted than Fig.~\ref{fig:knwonctf}e, which verifies the assertion that the early reflections provide more tolerance to the underestimation.
As will be shown in the following experiments, for the low input SNR case, the residual noise caused by the input noise will mask the musical noise. 
In addition, the residual noise can be mitigated by increasing the number of microphones.

\subsection{Experimental Results for the Noise-free Case}
For the proposed method and two competing methods, the quantitative performance measures, i.e. the averaged score over the 50 utterances, for the case of noise-free microphone signal are given in Table~\ref{tab:noise-free}. 
For each method, the performance measures for the unprocessed signal is computed based on the corresponding reference signal.
Thence the unprocessed signals have the different PESQ and LSD scores for the three methods. For all the three methods, it is not surprising that all the performance scores for the case of $T_{60}=0.79$ s are worse than the scores for the case of $T_{60}=0.5$ s.
In terms of SRMR, the proposed method achieves the highest scores among the three methods, with respect to the same baseline. This indicates that the reverberation residual of the proposed method is very low. 
For PESQ, the WPE method performs the best in terms of both the PESQ scores and the PESQ improvement (the increment from the PESQ score of the unprocessed signal to the score of the processed signal), and the proposed method outperforms the CDR method.  
The proposed method achieves the smallest LSD scores with the early reverberated signal as the reference. WPE has the highest LSD improvement with respect to the worst baseline.  
Overall, for the noise-free case, the proposed method and WPE achieves good performance measures, their outputs are fairly close to the early (64 ms) reverberated signal and the source signal, respectively.
As for the CDR method, the spectral magnitude of the diffuse reverberations cannot be accurately estimated using only the known DOA.

A sonogram example for the proposed method was shown in Fig.~\ref{fig:knwonctf}f, which intuitively illustrated the behavior of the proposed method. For comparison, Fig. \ref{fig:wpecdr} shows the sonogram examples for the WPE and CDR methods.
It can be seen that the output of WPE is well dereverberated and less distorted, while there are late reverberations remained in the output of CDR. 

\setlength{\tabcolsep}{4pt}
\begin{table}[t]
\centering
\caption{The performance measures of the three methods for the case of noise-free microphone signal. }
\label{tab:noise-free}
\begin{tabular}{c c | c c | c c | c c }  
	    &          		&\multicolumn{2}{c|}{WPE}    & \multicolumn{2}{c|}{CDR}      & \multicolumn{2}{c}{Proposed}  \\ \vspace{1mm}
	    &$T_{60}$ (s)      & unproc.  & proc.            & unproc.  & proc.               & unproc.   & proc.           \\ \hline
SRMR	    &0.5                & 2.42    & 3.23             & 2.42     & 2.61                & 2.42      & 3.56       \\ \vspace{2mm}
	    &0.79               & 1.93    & 3.12             & 1.93     & 2.34                & 1.93      & 3.32       \\  \hline  		    
PESQ	    &0.5                & 1.90    & 3.18             & 2.31     & 2.52                & 2.44      & 2.71   \\   \vspace{2mm}
	    &0.79               & 1.69    & 3.01             & 2.03     & 2.15                & 2.14      & 2.48   \\   \hline  
LSD 	    &0.5                & 8.28    & 3.37             & 4.49     & 4.00                & 3.95      & 2.99       \\ 
 	    &0.79               & 12.78   & 6.01             & 5.84     & 4.96               & 5.15      & 3.47       \\   
\end{tabular}
\vspace{-.3cm}
\end{table}

\subsection{Experimental Results for noisy case}  

\begin{figure*}[t]
\centering
\subfloat[SRMR $T_{60}=0.5$ s]{\includegraphics[width=0.32\textwidth,height=0.22\textwidth]{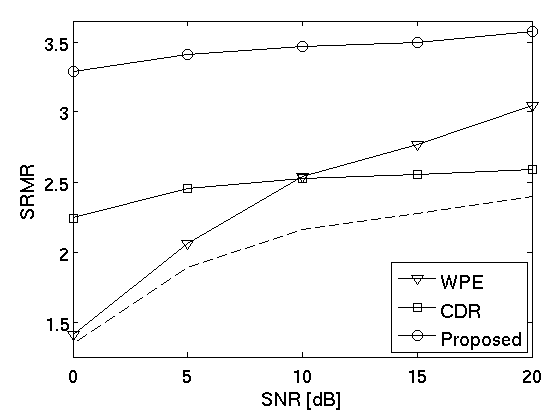}\label{subfig:na}}
\subfloat[PESQ $T_{60}=0.5$ s]{\includegraphics[width=0.32\textwidth,height=0.22\textwidth]{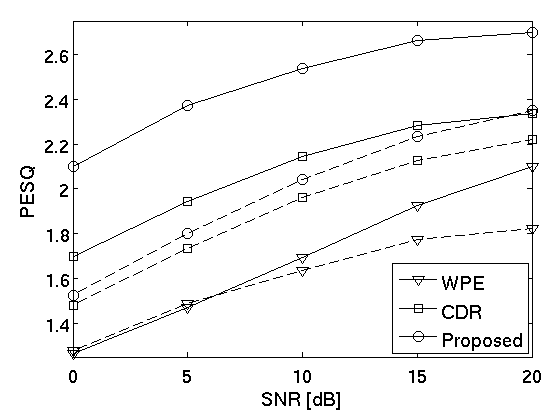}\label{subfig:nb}}
\subfloat[LSD $T_{60}=0.5$ s]{\includegraphics[width=0.32\textwidth,height=0.22\textwidth]{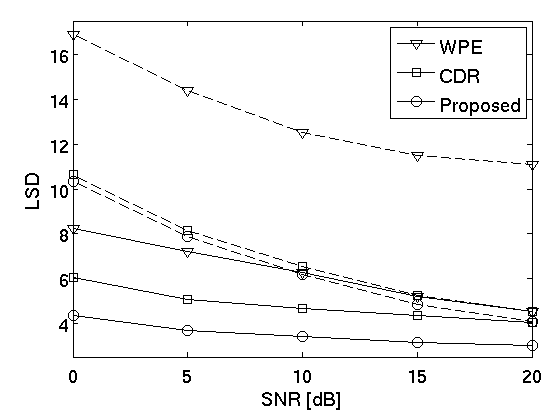}\label{subfig:nc}} \vspace{-.3cm}\\
\subfloat[SRMR $T_{60}=0.79$ s]{\includegraphics[width=0.32\textwidth,height=0.22\textwidth]{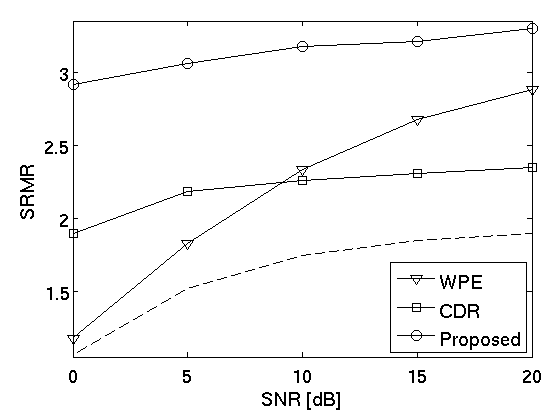}\label{subfig:nd}}
\subfloat[PESQ $T_{60}=0.79$ s]{\includegraphics[width=0.32\textwidth,height=0.22\textwidth]{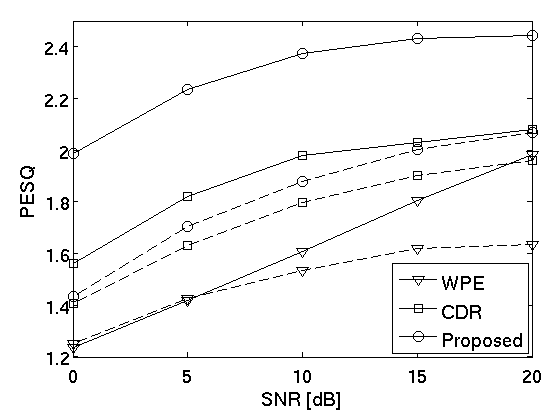}\label{subfig:ne}}
\subfloat[LSD $T_{60}=0.79$ s]{\includegraphics[width=0.32\textwidth,height=0.22\textwidth]{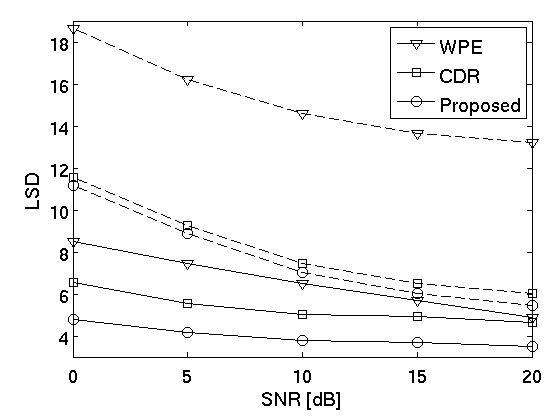}\label{subfig:nf}}
\caption{The performance measures for noisy signals as a function of SNR. The solid lines denote the scores of the processed signal for the three methods. 
The dashed lines with the same mark are the scores of the unprocessed signal taking the reference signal of the corresponding methods.} 
\label{fig:noisyresults}
\vspace{-0.3cm}
\end{figure*}

To evaluate the dereverberation and noise reduction capabilities of the proposed method, the noisy signals with various SNRs are tested. 
WPE is not designed for the noisy case, however, it is worthwhile to test its sensitivity to noise. Fig.~\ref{fig:noisyresults} depicts the performance measures of the three methods as a function of SNR. 
It can be seen that the performance measures for the cases of $T_{60}=0.5$ s and $T_{60}=0.79$ s are influenced by the noise signal in an identical way. 

From Fig.~\ref{subfig:na} and \ref{subfig:nd}, it can be seen that the SRMR score of the unprocessed signal decreases with the decreasing of SNR.  
The proposed method achieves the much higher SRMR scores than the other two methods. This confirms the efficiency of the proposed method for jointly dereverberation and noise reduction.
Along the decreasing of SNR, for the proposed method, the intense noise leads to more filter perturbations, noise residual and speech distortion.
For the CDR method, the noise leads to the larger variance of the CDR estimation, and consequently more noise residual. As a result,  the SRMR scores of CDR and the proposed method slowly decrease with the decreasing of SNR. 
However, WPE has a much larger degradation than the other two, which indicates the proposed method and CDR are less sensitivie to noise than WPE.
From Fig.~\ref{subfig:nb} and \ref{subfig:ne}, it is observed that the proposed method achieves the highest PESQ scores and PESQ improvement. 
The PESQ improvement of WPE is rapidly approaching to 0 with the decreasing of SNR.
From Fig.~\ref{subfig:ne} and \ref{subfig:nf}, it is seen that, the proposed method achieves the lowest LSD, which means that the outputs of the proposed method are close to the noise-free reference signal in terms of the logarithmic distance.

To intuitively illustrate the behaviors of three methods for noisy case, the sonogarm examples are given in Fig.~\ref{fig:noisy}. As seen,  WPE achieves a less reverberated speech signal than the microphone signal. 
However, it even amplifies the noise signal, as can be observed from the low frequency region where the WPE output is more noisy than the microphone signal. It is observed from Fig.~\ref{fig:noisy}c that the CDR output is less noisy than the WPE output. 
Fig.~\ref{fig:noisy}d shows that the output of the proposed method is much less noisy than the other two methods. This confirms that the proposed channel identification method (\ref{eq:ls}) is robust to the spatially uncorrelated microphone noise.
As metioned in Section~\ref{ssec:if}, the tolerance $\delta$ is set to be relatively small to avoid the signal distortion, thence we still can observe some residual noise, and the low power speech spectra  are still masked by the residual noise.
By informal listening test, the small replica and residual musical noise that present in the noise-free case are not clearly audible for this case, since the residual noise caused by the input noise is dominant, and this residual noise sounds close to the input noise. 

\section{Experiments with Real Recordings} \label{sec:experiments2}

To evaluate the proposed method in real world, and with multiple microphones,  we conducted two set of experiments using i) the multichannel impulse response data \cite{hadad2014}, and ii) the real recordings in REVERB challenge \cite{kinoshita2016}. 
The multichannel impulse response dataset is recorded using a 8-channel linear microphone array in the speech and acoustic lab of Ba Ilan University, with room size of $6$~m $\times$ $6$~m $\times$ $2.4$~m. The reverberation time is controlled by 60 panels covering the room facets.
The acoustic configuration of the impulse response dataset used in this experiment is  (i) reverberation time $T_{60}=0.61$~s, (ii) the microphone-to-source distance is 2 m, (iii) source direction is 
from $-90^{\circ}$ to $90^{\circ}$. The microphone array with 2-channel, 4-channel (the central two or four microphones) are tested for the proposed method.  
For each case, 50 TIMIT signals are convolved with the impulse responses with a random direction. The spatially uncorrelated stationary speech-like noise is added to the microphone signals with high SNR (20 dB) and low SNR (5 dB), respectively.
The RealData in REVERB challenge is recorded in a room with $T_{60}$ of 0.7 s. It contains 2 types of microphone-to-speaker distances, namely \emph{near} (1 m) and \emph{far} (2.5 m), which respectively have 90 and 89 recordings with different directions. 
We test the 2-channel and 8-channel RealData for development (dev). 

The same parameters are used as for the simulation dataset in Section \ref{sec:experiments1}, except that, according to the reverberation time, the filter length is approximately set to 320 ms, i.e. 20 taps, for both the two datasets.  
The noise PSD is estimated from the pure noise signals for the multichannel impulse response dataset, and is estimated using the noise PSD estimator \cite{mine2016assp} for the REVERB challenge dataset.

 \setlength{\tabcolsep}{3.0pt}
\begin{table}[t]
\centering
\caption{The performance measures for the multichannel impulse response dataset.}
\label{tab:mcird}
\begin{tabular}{c c | c c | c c c |  c c c   }   
	    &                   & \multicolumn{2}{c|}{CDR}  & \multicolumn{3}{c|}{WPE}      & \multicolumn{3}{c}{Proposed}  \\ \vspace{1mm}
	    &noise		& unproc.   & 2-ch     & unproc.   & 2-ch & 4-ch     & unproc.    & 2-ch & 4-ch  \\ \hline 
SRMR	    &20 dB               & 2.55     &  2.79    &2.55  & 3.05 & 3.12  & 2.55 &  3.47 & 3.48 \\ \vspace{2mm}
	    &5 dB               & 1.97    & 2.62   & 1.97 &  2.12  & 2.14  & 1.97    &3.25 & 3.32  \\  \hline  		    
PESQ	    &20 dB               & 2.55  &  2.83 & 2.00  &  2.36 & 2.42 & 2.67  &  2.92 &2.96\\   \vspace{2mm}
	    &5 dB               & 1.91  &  2.32 & 1.59  &  1.60 & 1.58&1.97   & 2.47 & 2.59 \\   \hline  
LSD 	    &20 dB               & 3.21 & 2.92  & 7.96 &3.50 &3.33 & 2.86  & 2.45 & 2.41  \\ 
 	    &5 dB               &      7.34 &3.78  &  11.48  & 6.50 &6.49  &  7.12 &  3.15 & 3.07 \\   
\end{tabular}
\vspace{-0.3cm}
\end{table}

The perfomance measures for the multichannel impulse response dataset are shown in Table~\ref{tab:mcird}. Note that CDR is only applicable for the  2-channel case, the filter length for WPE is set to 60 and 20 for the 2-channel and 4-channel cases, respectively. 
For the 2-channel case, it can be seen that the perfomance measures of all the three methods are almost consistent to the results obtained on the simulation data. 
The proposed method achieves better scores than the others. The WPE method performs worse for the low SNR case.
WPE achieves better scores with 4-channel than the 2-channel case for the high SNR case, while not for the low SNR case. 
The proposed method achieves better scores with the 4-channel than the 2-channel case for  both the two SNR cases. 
This indicates the efficiency of the multichannel extension of the proposed method. 
Both the channel identification method presented in Section~\ref{sec:ci} and the inverse filtering method presented in Section~\ref{sec:dereverberation} benifit from a larger number of microphones. 
For the channel identification method, the identification of each channel is carried out by using the cross-relations with all the other channels, thence a more robust identification can be achieved by increasing the number of channels. 
For the inverse filtering method, a larger number of channels will give a larger data size of the $\ell_2$ norm fitting problem in (\ref{eq:l2}) and (\ref{eq:l1}), which leads to a smaller error covariance of the least square estimation \cite{manolakis2005}.  
In addition, by listening test, the musical noise presented in the 2-channel case is noticeablely suppressed in the 4-channel case. 

The perfomance measures for the REVERB challenge dataset are shown in Table~\ref{tab:reverb}. 
CDR needs the prior knowledge of either the DOA or the noise coherence, or both, which are not available for this real recording case, thus CDR is not tested for this experiment. 
Only the SRMR score is given due to the lack of reference signal. The proposed method and WPE both require the prior knowledge of reverberation time. The filter length for WPE is set to 70 and 10 for the 2-channel and 8-channel cases, respectively.  
We can observe that the proposed method achieves better SRMR scores. In the dereverberated signal of WPE, the residual noise is noticeable.  
The \emph{near} case has a larger direct-to-reverberation ratio than the \emph{far} case, in other words, the desired direct-path signal (and early reverberations) is less contaminated by the late reverberations.
As a result, the SRMR scores of the \emph{near} case are higher.

The audio examples for all the datasets in this paper are available in our website.\footnote{https://team.inria.fr/perception/research/ctf-dereverberation}

 \setlength{\tabcolsep}{7.0pt}
\begin{table}[t]
\centering
\caption{The performance measures for the REVERB challenge dataset.}
\label{tab:reverb}
\begin{tabular}{c c | c  | c c  |  c c    }   
	    &                     &        & \multicolumn{2}{c|}{WPE}      & \multicolumn{2}{c}{Proposed}  \\ \vspace{1mm}
	    &dis		  & unproc.   & 2-ch & 8-ch         & 2-ch & 8-ch  \\ \hline 
SRMR	    &near                 &2.07       & 2.54 & 2.61         & 3.20 & 3.20 \\ \vspace{2mm}
	    &far                  &1.90       & 2.46  &2.54         & 3.03 & 3.07 		
\end{tabular}
\vspace{-0.5cm}
\end{table}

\section{Conclusion}\label{sec:conclusion}
In this paper, a blind multichannel speech dereverberation and noise reduction method has been proposed. 
The cross-relation method was extended to the STFT domain to circumvent the problem of near-common zeros for the long channel filters. 
The common zeros caused by the oversampling of STFT is solved by forcing the channel filters to perform the critical sampling. 
A constrained least-square problem was proposed to estimate the filters, which is robust to the noise interference and the filter length determination error. 
The channel filters were normalized by the first coefficient of one unique channel to remove the gain ambiguity accross subbands. 
The STFT domain inverse filtering can benifit from the sparsity of the source signal. An optimization problem with respect to the $\ell_1$ norm of the source signal and the $\ell_2$ norm fitting error between the microphone signals and the channel filtered  source signal 
was proposed to reduce the noise caused by both the filter perturbations and microphone noise interference.

A series of experiments using binaural data and multichannel data have been conducted. 
It is confirmed that the proposed channel identification method is efficient for the real RIRs, even for the high reverberant case. 
In addition, the channel identification method is robust to the spatially uncorrelated stationary noise, even for the low SNR case. 
In the proposed inverse filtering method, the automatically set tolerance for the $\ell_2$ norm fit works well for noise reduction. 
Overall, this paper proposes a practical real-world RIRs identification method, and thus a novel blind dereverberation and noise reduction method in the family of multichannel equalization technique.



\end{document}